# STRUCTURAL, ELECTRONIC PROPERTIES AND THE FEATURES OF CHEMICAL BONDING IN LAYERED 1111-OXYARSENIDES LaRhAsO AND LaIrAsO: *AB INITIO* MODELING

**V.V. Bannikov, I.R. Shein**

*Institute of Solid State Chemistry, Ural Branch Of Russian Academy of Sciences,
620990, Ekaterinburg, Pervomayskaya St., 91, Russia*

E-mail: bannikov@ihim.uran.ru

The comparative study of structural, electronic properties, topology of the Fermi surface, and the features of chemical bonding in layered 1111-oxyarsenides LaRhAsO and LaIrAsO has been performed based on the results of ab initio modeling of their electronic structure. It was established that only weak sensitivity with respect both to electron and hole doping is expected for LaIrAsO being non-magnetic metal, however, the Rh-containing compound should be characterized with weak band magnetism, and the hole doping is expected to be able to move its ground state away from the boundary of magnetic instability. The mentioned feature allows to consider LaRhAsO oxyarsenide as a possible "electron analogue" of LaFeAsO compound being the initial phase for the layered FeAs-superconductors.

INTRODUCTION

The oxychalcogenides and oxypnictides with common chemical formula $Me^{(a)}Me^{(b)}(Ch/Pn)O$ (where $Me^{(a)}$ is atom of Y, La, Bi or $4f$-metal, usually taking the oxidation state +3 in compounds, $Me^{(b)}$ is atom of $d$-metal, $Ch$ = S, Se, Te, $Pn$ = P, As, Sb) form the vast family of so-called layered 1111-phases [1,2] with tetragonal ZrCuSiAs-like structure which can be regarded as the sequence of $[Me^{(a)}$–O] and $[Me^{(b)}$–$(Ch/Pn)]$ blocks alternating along the tetragonal axis, where each atom is characterized with coordination number = 4 (Fig.1). Depending on the chemical composition, the 1111-phases can reveal the variations of a number of physical characteristics (such as electrical and thermal conductivity, some optical and magnetic properties, etc) in a wide range, and therefore they are of interest for various fields of modern materials science. For instance, as non-linear optic materials (LaCuSO, LaCuSeO) for various optoelectronic applications [3-7], materials for solar cells (LaZnPO:$Nd^{3+}$ [8]), ferro- and antiferromagnetic semiconductors (LaCu$_{1-x}$Mn$_x$SeO, LaMnPO [9-11]) or even the solids with non-collinear magnetic structure (some compounds with 4*f*-metals), thermoelectric materials (BiCuSO, BiCuSeO [12,13]) and so on. Besides, some non-magnetic 1111-phases can be proposed as the initial crystalline matrices for design of novel magnetic materials with abilities of "fine tuning" of their functional properties by means of the corresponding choice of dopants (see [14] and References therein). But the most significant interest to 1111-oxypnictides is due to the superconductivity discovered in 2008 in LaFeAsO$_{1-x}$F$_x$ system ($T_c$=26 K [15,16]), as well as in some other layered [FeAs]-phases, for instance, in 111-systems LiFeAs [17] and NaFeAs [18], in 122-phase SrFe$_2$As$_2$ doped with cobalt [19,20], in 21311-system Sr$_4$V$_2$O$_6$Fe$_2$As$_2$ [21] and so on. This fact is interesting in that the diamagnetism of compounds being rich in transition metal atoms which is associated with superconductivity seems to be quite unexpected. Nevertheless, the transition to superconducting state was also observed in a number of other 1111-phases including atoms with open 3*d*- or 4*f*-shell, for example, in LaNiPO [22], LaNiAsO$_{1-x}$F$_x$ [23], in *RE*FeAsO systems (*RE* = Ce, Pr, Nd, Sm [24]) doped with cobalt, in non-stoichiometric BiCuSO [25], etc. So, 1111-oxypnictides and oxyhalcogenides can be treated as a promising class of compounds to discover new superconducting materials with high $T_c$ values.

Let us emphasize that the most of the known 1111-superconductors are *doped* compounds, at the same time, they are less common among the pristine phases, and their critical temperature $T_c$ is relatively small, for example, LaFePO (4 K), LaNiPO (3 K), LaNiAsO (2.4K) [26]. In turn, the pure LaFeAsO compound is an antiferromagnetic semiconductor that undergoes a structural phase transition at T <150 K, associated with turning into so-called SDW (*spin density wave*) state [16,26], however, the fluorine doping on the oxygen sublattice (of *electronic* type) suppresses the mentioned transitions, resulting in move away from the boundary of magnetic instability of the system with its further turning to superconducting state at T <$T_c$. As an alternative strategy of electron doping, the introduction of bivalent strontium into the lanthanum sublattice [27] can be proposed, for example, as well as the partial substitution for iron of *d*-metal atoms having a higher number of valence electrons. It was reported in [28] about the successful synthesis of LaFe$_{1-x}$Rh$_x$AsO solid solutions for a wide range of concentrations x=0.05–0.75, which reveal superconductivity at x=0.05–0.15 (T$_c$ ~ 12–15 K), as well as about a new 1111-phase LaRhAsO being a common (non-superconducting) metal at temperatures down to 5 K, and later the information on the synthesis and the structural properties of the isoelectronic 1111-oxyarsenide LaIrAsO (as well as on *Ln*(Rh/Ir)AsO phases, where *Ln* = Ce, Nd) has been published [29]. The atom of Rh/Ir has one valence electron more than the iron atom, therefore, the higher values of electron concentration are expected for La(Rh/Ir)AsO systems, as compared to LaFeAsO, and it can be assumed that the inverse situation could occur for them: i.e. a hypothetical transition to the superconducting state could occur due to *hole* doping (for example, with nitrogen implanted into oxygen sublattice). However, to estimate the validity of such an assumption the detailed information about the features of the band structure and chemical bonding in LaRhAsO and LaIrAsO phases is necessary, while all available data on the properties of these systems (to the best of authors knowledge) are currently confined to the two papers mentioned above, in which the problems concerning to their electronic structure were not considered.

The goal of this paper is to replenish this lack of information. Based on the results of *ab initio* band calculations, we carried out a comparative study of the structural properties, electronic structure, topology of the Fermi surface, as well as the features of chemical bonding in the isostructural and isoelectronic 1111-phases LaRhAsO and LaIrAsO. Taking into account the peculiarities of their band structure we have made the conclusions about their relative sensitivity to electron and hole doping, and also about its possible influence on their electronic and magnetic properties.

THE MODEL AND THE COMPUTATIONAL METHOD

LaRhAsO and LaIrAsO oxyarsenides are characterized with tetragonal structure of ZrCuSiAs type (space group $P4/nmm$, №. 129), composed, as mentioned above, of alternating blocks …[LaO]/[(Rh/Ir)As]/[LaO]/[(Rh/Ir)As]/[LaO]… along the tetragonal axis. The positions of atoms in the unit cell containing two formula units are: La (1/4, 1/4, $z_{La}$), Rh/Ir (3/4, 1/4, 1/2), As (1/4, 1/4, $z_{As}$) and O (3/4, 1/4, 0), where $z_{La}$ and $z_{As}$ are the *internal coordinates* of atoms, whose values remain undefined, if we would base only on symmetry considerations. In the papers [28–29] only the values of lattice constants $a$ and $c$ for LaRhAsO and LaIrAsO were provided, but not those of $z_{La}$ and $z_{As}$, so, at the initial stage of the study, the crystal structure of oxyarsenides was fully optimized – both the lattice constants and the internal coordinates, and those optimized values were employed for further calculations.

The electronic structure of the compounds under consideration was modeled using the full-potential linearized augmented plane wave method (FP-LAPW) [30], implemented in Wien2k software package [31], within the generalized gradient approximation (GGA) of exchange-correlation potential in PBE form [32]. The densities of electronic states (DOS) were calculated within the modified tetrahedron method [33]. The radii of atomic *muffin-tin* (MT) spheres were chosen to be 2.2, 2.0 and 1.8 Bohr for La/Rh/Ir, As and и O, respectively. The employed plane wave basis set was confined by the value of $K_{max}$ defined from the relation $R_{min} \times K_{max} = 7$, where Rmin is the minimal MT radius. The value of $E_{cut-off}$ energy separating the valence states from the core ones was taken to be –6.0 Ry. The integration over the Brillouin zone (BZ) was performed using 12×12×6 *k*-mesh (63 *k*-points in the irreducible part of BZ), that is sufficient for modeling the electronic spectrum of the considered compounds with reasonable accuracy, however, for the investigations of topology of the Fermi surface much more dense *k*-mesh was required (about 3000 *k*-points in the irreducible part of BZ). The convergence criteria of self-consistent calculations was set to be 0.00005 Ry for total energy, 0.001 $e$ for electronic charges within MT spheres, and 1 mRy/Bohr for atomic forces, – as calculated at consecutive steps.

RESULTS AND DISCUSSION

The optimized values of structural constants and internal coordinates for LaRhAsO and LaIrAsO in comparison with the available experimental data are given in the Table. For the Rh-containing phase, the calculated value of the lattice constant $a$ is approximately 0.016 Å more than that for the Ir-containing compound is; at the same time, the value of the constant $c$ is approximately 0.131 Å less. However, such significant (at first sight) difference in the values of the constant $c$ (taking into account that the atomic radii of Rh and Ir are very close - 1.34 and 1.35 Å, respectively) seems quite reasonable if we compare the interatomic distances in these crystals. Indeed, the distances (Rh-As) and (Ir-As) are 2.464 Å and 2.479 Å, respectively, and their difference is comparable to that in the atomic radii of Rh and Ir; the interatomic distances (La-O) in LaRhAsO and LaIrAsO differ relatively little – 2.396 Å and 2.402 Å, and the distances between the nearest La and As atoms are 3.417 Å and 3.409 Å, respectively. In other words, an increase in the parameter $c$ (as Ir is substituted for Rh) changes the interatomic distances not too significantly (~0.01 Å or less), at the same time, the difference in the angles of Rh-As-Rh and Ir-As-Ir bonds is about 1° (72.96° and 72.12°, respectively). It should be noted that the angle of Fe-As/Se-Fe bonds is an important characteristic of layered (Fe-As/Se) superconductors, in particular, for the systems based on 122-phases (such as $CsFe_2Se_2$, $KFe_2As_2$) the correlation between its value and the temperature of the superconducting transition, the value of electron-phonon coupling constant, etc can be pointed out [34, 35]. The relative deviation of the calculated value of the parameter $a$ from its experimental value, $|a^{(calc)}/a^{(exp)} – 1|\times100\%$, for LaRhAsO and LaIrAsO is 0.46% and 0.95%, and that for the parameter $c$ is 1.48% and 1.97%, respectively. The relatively large value of the latter is probably due to the fact that the value of the parameter *c* is determined mainly by the interactions between the structural blocks of the layered phase, in which the van der Waals contribution typically ignored in DFT calculations can play a certain role. We note that a comparable relative deviation of the parameter *c* (~1.5–2%) from its experimental value also takes place for other 1111-phases, for example, for systems of the family (Sr/Ba)Ag*Ch*F (*Ch* = S, Se, Te) [36].

Fig. 2 presents the total and partial DOS for the investigated oxyarsenides. The structure and the composition of the spectrum of the valence states of LaIrAsO can be characterized as follows. It consists of five main bands - A, B, C, D, E (see overview spectrum in the inset, Fig. 2b). The band A is located in the range of about 14.5–16.8 eV below the Fermi level ($E_F$) and is composed mainly of La-5$p$ states, to which 2$s$ states of oxygen are mixed. The band B is separated from it by a gap about 1.25 eV and is composed of As-4$s$ states with a small admixture of Ir-5$d$. The partially filled valence band C is located in the range from about 7.4 eV below $E_F$ to 1.7 eV above $E_F$, and its composition is most complex: Ir-5$d$ and As-4$p$ states contribute to its entire energy range, but their role is dominant at the bottom and in the upper part of the C band, while in its middle part (3–6 eV below $E_F$) the contribution of the O-2$p$ states (with an admixture of La-5$p$, 5$d$ states) also is significant, in other intervals it is relatively small. The vacant band D represents a sharp peak located in the range of about 1.7–3.5 eV above the $E_F$ and is composed of 4$f$ states of lanthanum, the E band is adjacent to it and is formed mainly by La-5$d$ states with a small admixture of the states of all other atoms. The structure and composition of the LaRhAsO valence spectrum do not differ qualitatively, it is only essential to note that the filled part of the valence band C for the Rh-containing phase is narrower and its width is approximately 6.1 eV. It follows from the character of band spectrum that both of these compounds should be characterized by metallic conductivity, – in accordance with the experimental data [28-29], further, it can be seen from Fig.2 that in the vicinity of $E_F$ the contribution of the [La-O] blocks to the total DOS is negligible as compared to that of the [(Rh/Ir)-As] blocks, so, the layered 1111-phases La(Rh/Ir)AsO can be considered as alternating "conducting" [(Rh/Ir)-As] blocks separated by [La-O] "insulating gaskets".

The most significant difference in the electronic structure of the 1111-phases under consideration takes place in the close vicinity of $E_F$. The upper part of the valence band C has a complex multi-peak structure, for LaIrAsO $E_F$ is placed on the front of the lower peak, while for LaRhAsO it almost coincides with peak maximum. For further clarity, Fig. 3 shows the character of the dispersion of the energy bands for the investigated oxyarsenides. LaIrAsO is characterized by a relatively simple $E(\mathbf{k})$ dependence in the vicinity of $E_F$, which includes only highly dispersed bands, however, for LaRhAsO the picture becomes more complicated: the additional $E(\mathbf{k})$ branches appear which correspond to Γ-A and Γ-M directions in BZ, and which reveal low-dispersion behavior near $E_F$ (quasi-flat zones) corresponding to the maximum of the DOS peak at $E_F$, in addition, two highly dispersed branches $E(\mathbf{k})$ are shifted to the vicinity of $E_F$ - with a minimum at Γ point, placed about 0.4 eV below $E_F$, and with a minimum at point A, placed almost at $E_F$. For LaIrAsO, the mentioned bands are placed above the $E_F$ and do not affect its electronic structure in the ground state. It is known that the DOS maximum at $E_F$ favors the stabilization of the ferromagnetic state of the system, in the simplest case this is expressed by well-known Stoner criterion: $D(E_F) \cdot J > 1$, where $J$ is the exchange parameter, and $D(E_F)$ is the DOS value at the Fermi level. Actually, the spin-polarized calculations of LaRhAsO band structure indicate the presence of exchange splitting of the near-Fermi states (see inset in Fig. 2a), as a result, the small magnetic moments (~0.06 $\mu_B$) are induced on rhodium atoms, and the DOS values at $E_F$ for the electronic sub-systems "*spin up*" and "*spin down*" become different, $D_\uparrow(E_F)=3.97$ states/eV and $D_\downarrow(E_F)=1.91$ states/eV, respectively, and the magnitude of spin polarization, $SP=|D_\uparrow(E_F)–D_\downarrow(E_F)|/|D_\uparrow(E_F)+D_\downarrow(E_F)|$, is approximately 35%. The mentioned circumstance allows to characterize LaRhAsO as a substance with weak band magnetism or, in other words, with exchange-enhanced Pauli paramagnetism [37]. At the same time, for LaIrAsO, the exchange splitting of states is insignificant, and this compound can be characterized as an ordinary nonmagnetic metal. The mentioned differences in the band structure of these isoelectronic phases can presumably be attributed to the difference in the As-Rh/Ir-As bond angles (see above), resulting in a different overlap measure of the *p*, *d*-orbitals of the Rh/Ir and arsenic atoms. Taking into account the considered features of the electronic structure, we can assume that LaIrAsO should be relatively insensitive to both electronic and hole doping (when low concentrations of the dopant change the character of the electronic spectrum just slightly, and a rigid band model can be reliable), – actually, it would result just in $E_F$ shifts along the front of the lower DOS peak in the near-Fermi region, however, the electronic properties of LaRhAsO can be expected to depend critically both on the concentration and the type of dopant. It was established [28] that the fluorine-doped system $LaRhAsO_{0.87}F_{0.13}$ does not reveal superconductivity down to 5 K, and its electrical resistance is slightly lower than that of the LaRhAsO pristine phase. It becomes clear from the DOS distribution near $E_F$ (Fig. 2a) that doping with fluorine (*electron* doping) should simply result in a shift of the Fermi level to higher energies with a relatively small variation of $D(E_F)$, as well as in increase in the concentration of conduction electrons. At the same time, *hole* doping and the corresponding shift of $E_F$ downward could result in an abrupt fall in the value of $D(E_F)$, in other words, to movement away from the boundary of the magnetic instability of the system and to the stabilization of its non-magnetic state.

Of course, the discrepancies in the band structure of compounds in the close vicinity of $E_F$ result in different topology of their Fermi surface (FS), shown for LaRhAsO and LaIrAsO in Fig.4. For the Rh-containing phase, the important feature of its FS is the quasi-2D character with respect to Γ-Z direction in BZ (i.e. its cross section in the perpendicular plane remains almost unchanged along the indicated direction in the entire BZ), while for the Ir-containing phase, the FS does not possess this property. A common trait of the FS for both systems is the presence of bent quasi-2D sheets at the corners of BZ, which correspond to the intersection of $E_F$ with highly dispersed $E(\mathbf{k})$ branches for Γ-A, A-X and Γ-M directions, and separate the filled electronic states (in BZ corners) from the vacant ones. For LaIrAsO, the states with $\mathbf{k}$-vectors from the vicinity of Γ point are vacant, but the "electronic pockets" corresponding to the vicinity of Z point are present, while for LaRhAsO the situation is principally different: one of the FS sheets represents an axial "tube" oriented along Γ-Z direction, and the points of the $\mathbf{k}$-space being inside it correspond to the filled electronic states. It is known that the FS of oxypnictides LaFeAsO and LaFePO have a similar feature, but in that case these "tubes" are of *hole*-type [38], and this discrepancy is obviously associated with different electron concentration in the systems. We also note that for LaRhAsO, the presence of quasi-flat bands $E(\mathbf{k})$ in the close vicinity of $E_F$ results in appearance of additional FS sheets in the vicinity of the BZ vertices. However, the volume of $\mathbf{k}$-space confined by them is negligible, and a discussion of their topology and the corresponding electronic properties of this compound is beyond the frameworks of this paper.

It is convenient to illustrate the specific features of the chemical bonding in layered 1111-oxyarsenides with the maps of charge density distribution in the crystallographic plane (400). In Fig. 5 they are shown for LaRhAsO, the character of the charge distribution for the Ir-containing phase is almost identical. The overall chemical bonding picture in the considered compounds is typical for the layered 1111-phases in general - inside [La-O] and [Rh-As] structural blocks, the bonding is of mixed covalent-ionic type (in Fig. 5a the formation of the corresponding directional bonds is visible), while there are no directional bonds between the atoms from different structural blocks, and their interaction is of ionic type. Indeed, [La-O] blocks play the role of "charge donors" in the crystal, the estimated charge transfer

[La-O]$^{\delta+}$→[Rh-As]$^{\delta-}$ between the structural blocks is approximately 0.62 *e*, resulting in electrostatic interaction between them. The distribution of charge density shown in Fig. 5a has been obtained as *all* valence electron states for LaRhAsO were taken into account, while for more clarity, Fig. 5b shows the distribution calculated as *only* the contributions of the states of the partially filled band C (Fig. 2) were included. Taking into account the its partial composition, it can be argued that the states responsible for the formation of the covalent component of La-O and Rh-As bonds are (La-5*p*, -5*d*, O-2*p*) and (Ir-5*d*, As-4*p*), respectively. The states of the lower bands A and B practically do not contribute to the charge density shared by atoms, being mostly localized. We note that for Rh-As bonds, the maximum of the distribution of the inter-atomic charge density is not oriented along the line connecting the atomic cores, but rather "wraps" the arsenic atoms (Fig.5b), this feature can be explained both by geometry of the structural blocks (the network of distorted Rh-As tetrahedra) and the type of atomic orbitals (5*d*, 4*p*) involved in the formation of bonds. One more peculiarity worth to mention is the formation of direct As-As inter-atomic bonds in 1111-oxyarsenides. It is seen from Fig. 5c that for LaRhAsO and LaIrAsO (c1 and c2, respectively), the character of charge density distribution in (110) plane between the nearest arsenic atoms is typical for directional covalent bonds, however, in As-planes of the crystal (parallel to (100) plane) directional bonds between arsenic atoms are absent.

CONCLUSION

Based on the results of the first-principles band structure calculations, a comparative study of the structural, electronic properties, as well as features of the chemical bonding in the layered 1111-phases of LaRhAsO and LaIrAsO was performed. Both compounds are characterized with a metallic-like electronic spectrum, and they can be regarded as alternating "conducting" [(Rh/Ir)-As] blocks separated by [La-O] "insulating gaskets". The most essential difference in their band structure is revealed in the close vicinity of $E_F$: while for LaIrAsO there only highly dispersed branches E(*k*) are present, for LaRhAsO it includes additional quasi-flat bands. That results in the appearance of DOS peak at $E_F$ and, as a consequence, the stabilization of a weak band magnetism of the Rh-containing phase, while LaIrAsO is a conventional non-magnetic metal. As an important feature of the Fermi surface topology for LaRhAsO its quasi-2D character can be pointed out, as well as the presence of axial "electronic tube" oriented along the Γ-Z direction of the Brillouin zone (by analogy with the "hole-type tubes" of Fermi surface for LaFeAsO and LaFePO). The chemical bonding inside the [La-O] and [(Rh/Ir)-As] structural blocks is of mixed covalent-ionic type, while the bonding between atoms from different structural blocks is of ionic type due to charge transfer $[La-O]^{\delta+} \rightarrow [Rh-As]^{\delta-}$ between them, its other important peculiarity is the formation of direct As-As covalent bonds in LaRhAsO and LaIrAsO. Taking into account the mentioned features of the electronic structure, it can be assumed that LaIrAsO compound should be relatively insensitive to both electronic and hole doping, while for LaRhAsO, the hole doping could result in a movement away from the magnetic instability boundary of the system and stabilize its non-magnetic state, – by analogy with electronic doping of LaFeAsO resulting in stabilization of its diamagnetic state and superconductivity.

ACKNOWLEDGEMENT

THE TABLE

Optimized values of the lattice constants $a$, $c$ (in Å) and internal coordinates $z_{La}$, $z_{As}$ for La(Rh/Ir)AsO oxyarsenides in comparison with available experimental data (a − [28], b − [29] ).

|  | $a$ | $c$ | $c/a$ | $z_{La}$ | $z_{As}$ |
|---|---|---|---|---|---|
| LaRhAsO | 4.142 (4.123)[a] | 8.594 (8.469)[a] | 2.0748 | 0.14016 | 0.65517 |
| LaIrAsO | 4.126 (4.087)[b] | 8.725 (8.557)[b] | 2.1146 | 0.14077 | 0.65742 |

FIGURES

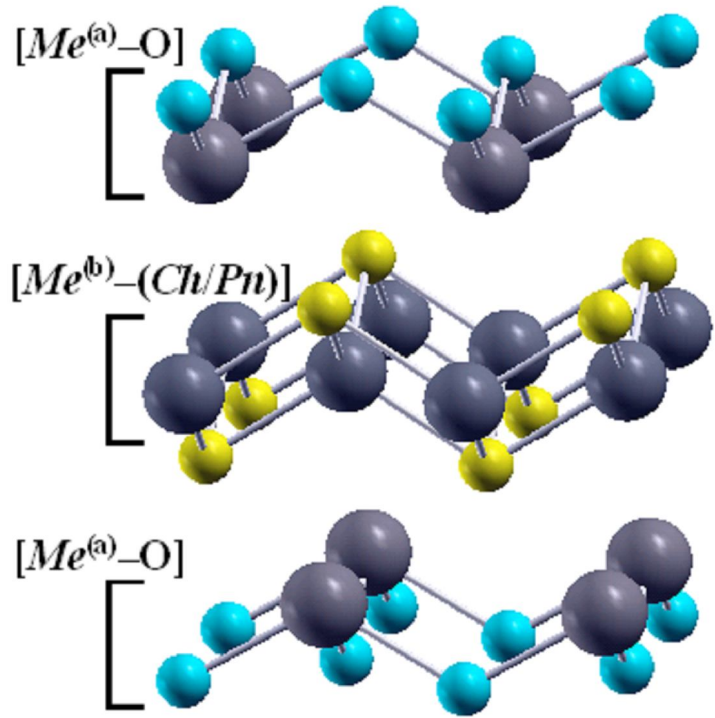

Fig.1. A fragment of the crystal structure of layered 1111-phases with common chemical formula $Me^{(a)}Me^{(b)}(Ch/Pn)O$ (see text). The yellow balls are the $Ch/Pn$ atoms, the light blue ones are the oxygen atoms, and the gray balls are the $Me^{(a)}$ and $Me^{(b)}$ atoms in the corresponding structural blocks.

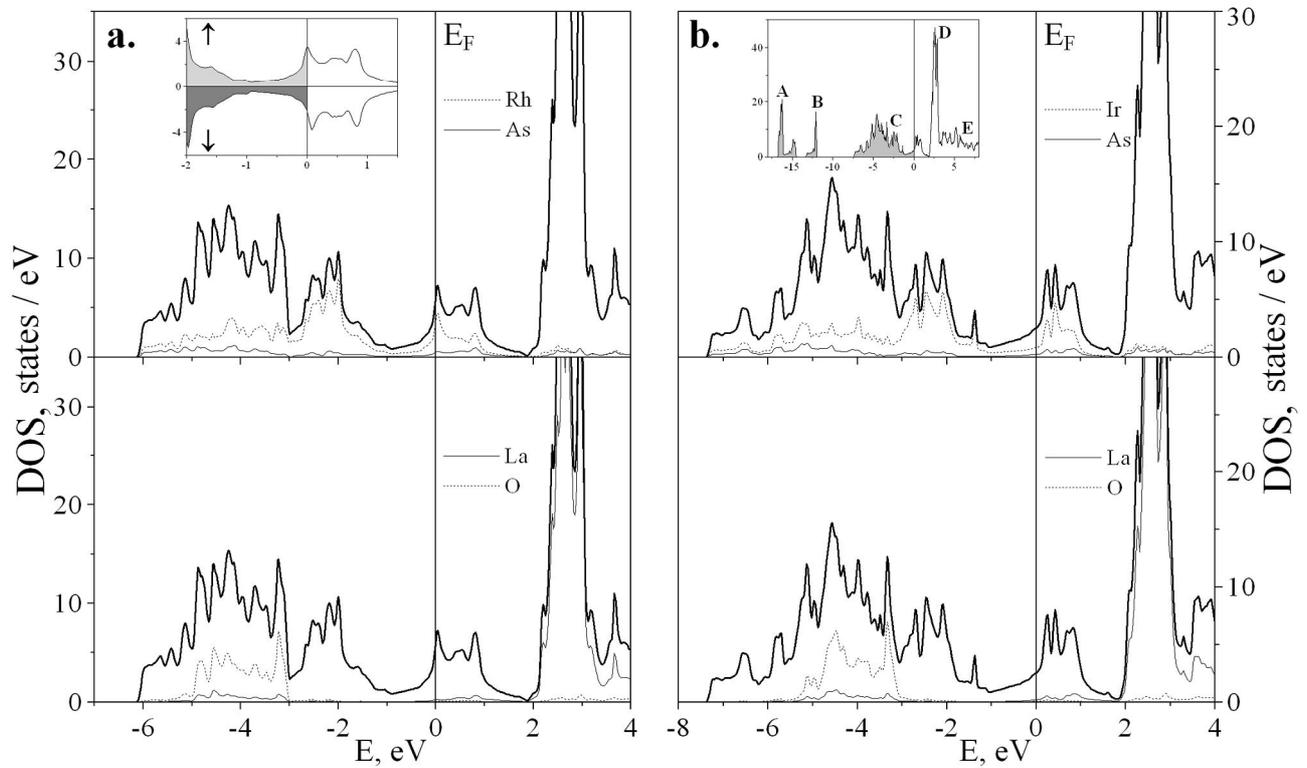

Fig.2. Total and partial DOS for 1111-oxyarsenides LaRhAsO (a) and LaIrAsO (b). Total DOS are shown with bold curves on all plots. Insets: (a) – the character of spin polarization of near-Fermi states for LaRhAsO, (b) – an overview spectrum of the LaIrAsO valence states.

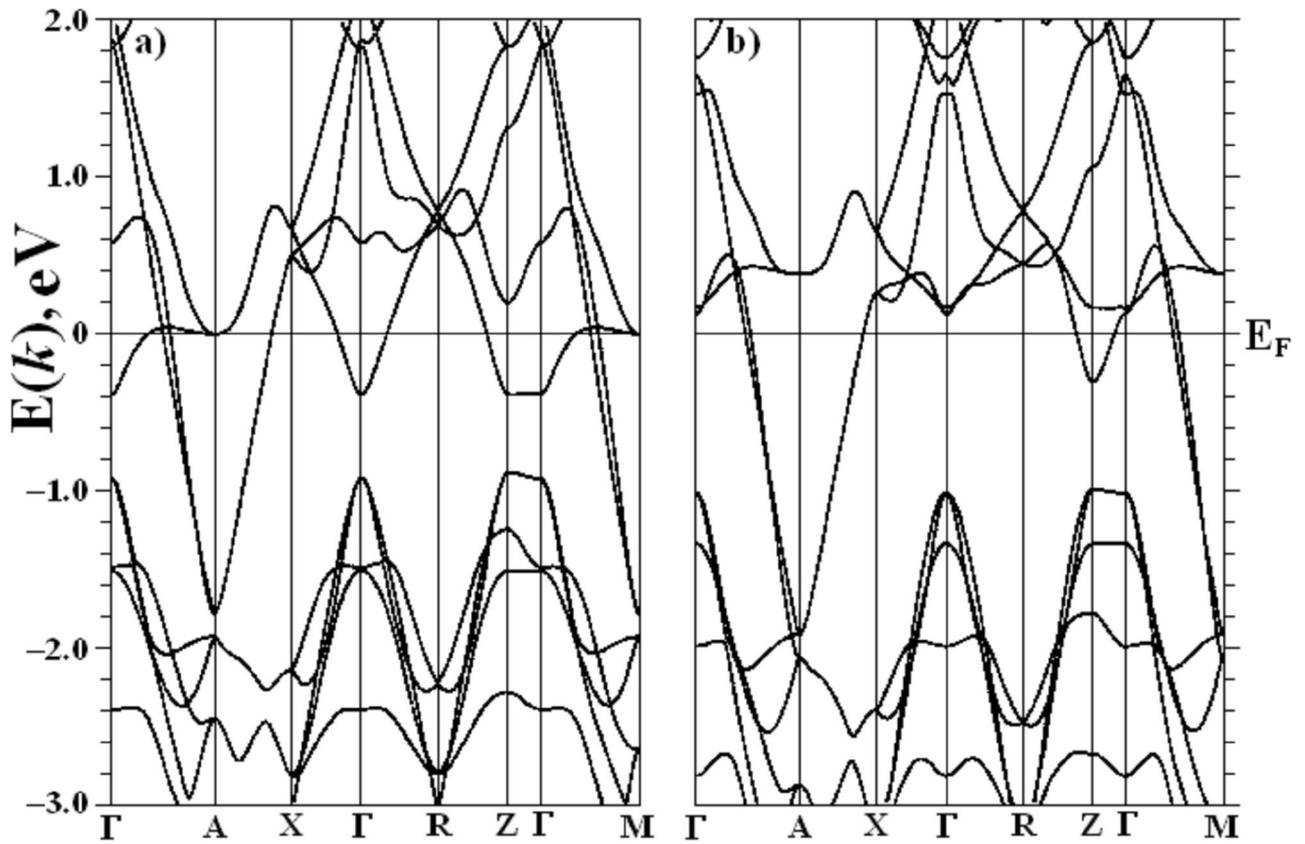

Fig.3. The character of the energy bands dispersion, E(*k*) for LaRhAsO (a) and LaIrAsO (b).

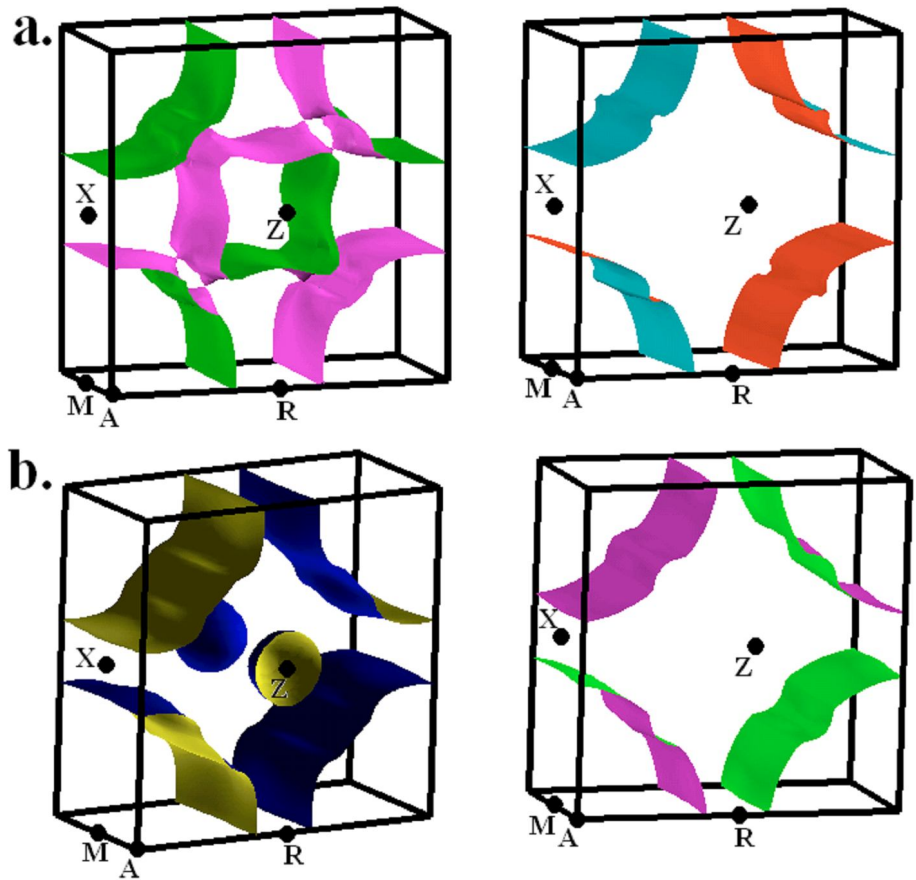

Fig.4. The Fermi surface for LaRhAsO (a) and LaIrAsO (b). For more clarity, the constituent sheets are shown separately and at different view angles; Γ point at the center of the Brillouin zone is not marked, but implied.

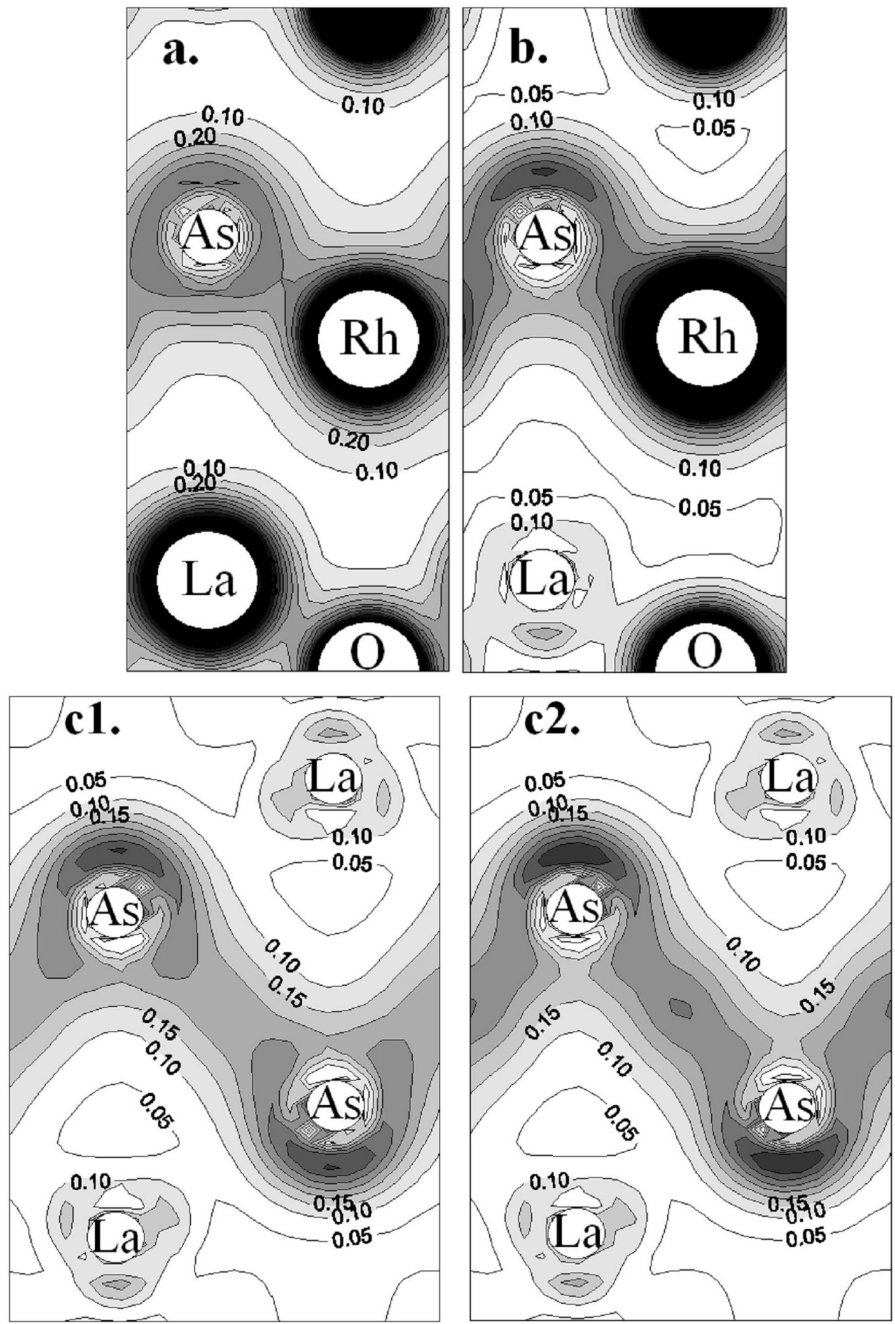

Fig.5. The maps of charge density distribution: a, b – in (400) plane for LaRhAsO (a – the contributions of all valence states were taken into account, b – *only* the contributions of the states of C band, see Fig.2); c1, c2 – in (110) plane for LaRhAsO and LaIrAsO, respectively (only the contributions of C band were taken into account). The values corresponding to the isoelectronic lines are in units of $e/\text{Å}^3$.